\documentclass[conference]{IEEEtran}

\newcommand{\sv}{{\bf \bar{ s}} }
\newcommand{\rhov}{ \hat{ \rho } }
\newcommand{\Pv}{{ \bf P} }

\usepackage{graphicx}
\usepackage{amsfonts}
\usepackage{amssymb}
\usepackage{amsmath}
\usepackage{float}

\title{A Flow Level Perspective on Base Station Power Allocation in Green Networks}

\author {Vineeth S Varma$^{1,2}$, Salah Eddine Elayoubi$^1$, Samson Lasaulce$^2$ and Merouane Debbah$^3$ \fontsize{9}{8} \\\\ \begin{minipage}{131pt} \centering $^{1}$Orange Labs\\92130 Issy Les Moulineaux\\France\\vineeth.svarma@orange.com\\salaheddine.elayoubi@orange.com
\end{minipage}
\begin{minipage}{0pt}
\end{minipage}
\begin{minipage}{130pt}\centering  $^{2}$LSS, SUPELEC\\ 91192 Gif sur Yvette\\ France\\
Samson.lasaulce@lss.supelec.fr\\

\end{minipage}
\begin{minipage}{30pt}  \space
\end{minipage}
\begin{minipage}{120pt}\centering  $^{3}$Alcatel Lucent Chair\\SUPELEC\\ 91192 Gif sur Yvette\\ France\\
Merouane.debbah@supelec.fr
\end{minipage}}

\begin{document}
\maketitle

\begin{abstract}

In this work, we propose a novel power allocation mechanism which allows one to optimize the energy-efficiency of base stations operating in the downlink. The energy-efficiency refers to the amount of bits that can be transmitted by the base station per unit of energy consumed. This work studies the impact of flow-level dynamics on the energy efficiency of base stations, by considering user arrivals and departures. Our proposed power allocation scheme optimizes the energy-efficiency, accounting for the dynamic nature of users (referred to as the global energy-efficiency). We emphasize our numerical results that study the influence of the radio conditions, transmit power and the user traffic on the energy-efficiency in an LTE compliant framework. Finally, we show that the power allocation scheme that considers traffic dynamics, is significantly different from the power allocation scheme when the number of users is considered as constant, and that it has a better performance.

\end{abstract}

\section{Introduction}

For a long time, the problem of energy in the field of communications revolved around autonomous, embarked, or mobile terminals. Nowadays, with the existence of large networks involving both fixed and nomadic terminals and the larger data rates supported, the energy consumed by the fixed infrastructure has also become a central issue for communications engineers \cite{salah}. As stated by the project Green-Touch, the telecommunications industry currently account for $2\%$ of the global carbon footprint, of which the major portion comes through the energy consumed at base stations \cite{greentouch}. This has led to the growing awareness for the need to reduce energy consumption as well as to optimize the use of energy in order to gain maximum benefit out of every unit of energy spent. The present work falls into this framework, more specifically, our goal is to devise the power allocation schemes for base stations in green wireless networks with the focus on downlink. The novelty of this work is in treating the problem of energy-efficiency and power allocation for dynamic users, i.e for users who, like in most practical cases, arrive randomly with a finite workload and depart after finishing it.

Among the pioneering works on energy-efficient power control is the work by Goodman et al \cite{goodman} in which the authors define the energy-efficiency of a communication as the ratio of the net data rate to the radiated power; the corresponding quantity is a measure of the average number of bits successfully received per joule consumed at the transmitter. This metric has motivated many works.  A survey on works that deal with this metric can be found in \cite{veronica-survey}. Other works like \cite{meshkatiCDMA06} deal with the energy-efficiency metric, and it is applied to the problem of distributed power allocation in multi-carrier CDMA (code division multiple access systems) systems, in \cite{Mesh09} it is used to model the users delay requirements in energy-efficient systems.

Summarizing the literature overview for energy-efficiency optimization, we conclude that although several works consider deal with this problem, they do not take into account several key-aspects of the network. First, in the definition of energy-efficiency, the number of users in the system is fixed, corresponding to a full buffer traffic model. In a real system, users arrive and depart and the number of users in the system is a dynamic quantity. Secondly, the transmission cost usually corresponds to the radiated power that is, the power of the radio-frequency signals. In this paper, we propose a power allocation scheme that responds to these two needs: considering the dynamic behaviour of users and taking into account the whole power consumption and not only the radiated power. This work uses a cross-layer approach, which deals with both the Media Access Control (MAC) layer, as well as the flow level (user arrivals and departures) in Orthogonal Frequency-Division Multiple Access (OFDMA) systems that are LTE compliant. Similar cross-layer approaches have been used in works like \cite{AltmanCons07} and \cite{richard}, but the metric used is often the capacity or data rates maximized under power constraints, while in this work we deal here with energy-efficiency optimization.

The original contributions of this paper are summarized as follows:
\begin{enumerate}
\item We consider a new energy efficiency metric that accounts for the overall power consumption of the base station, including common channel and fixed consumption parts.
\item We derive an optimal power allocation scheme that maximizes the energy efficiency, while preserving Quality of Service (QoS).
\item We show that the power allocation that considers the dynamic behavior of users is significantly different from the scheme optimized locally for each state of the network. In addition to that, the former performs better than the latter. To the best of our knowledge, this is the first time where such a flow level power allocation scheme is derived.
\end{enumerate}
This paper is structured as follows. In Section II, we present the system model and define the proposed performance metric. In Section III, we derive the optimal power allocation scheme when supposing that the number of users is fixed. Section IV shows how to deal with the dynamic behavior of users. Section V provides numerical results comparing both approaches (local vs. global optimization). Finally, we conclude the paper and suggest some possible extensions to this work.

\section{System model}

\subsection{System description}

We consider a transmitting base station with buffers of infinite (or very large) size. The base station sends packets into a queue for each user which is stored in these buffers. The packets arrive at each time slot $T_P$ (expressed in seconds), each packet being of size $S_p$ (expressed in bits). The data rate $R_p$ is equal to $\frac{S_p}{T_P}$. The throughput when using all the available bandwidth is denoted by $R(\rho)$ (expressed in bits per second), when the receiver has an average signal to interference plus noise ratio (SINR) of $\rho$. This SINR depends directly on the transmit power $P$ (expressed in Watts) as $\rho = \frac{P}{\sigma^2}$. Here $\sigma^2$ represents the average noise for a given radio condition (expressed in Watts) and it depends on the distance of the receiver from the base station.  Note that in this work, the effects of fast fading are not studied and we just consider the average SINR.

All packets of a user are assumed of the same size and the average throughput on the radio interface, when the queue for the corresponding user is active, is denoted by $R_a(\rho)$ (expressed in bits per second) which depends on the bandwidth available. When all the packets in the queue are transmitted the queue becomes empty and inactive. We assume that the transmitter always transmits packets while the queue is not empty. Each packet stored in the buffer is a collection of frames that are transmitted over the symbol time $T_s$ (expressed in seconds). Each frame is transmitted or retransmitted till it goes through and an acknowledgment is received. With these assumptions we proceed to calculate the average packet duration $T_d$ in the buffer.
\begin{equation}
T_d = \frac{S_p }{ R_a(\rho)}
\end{equation}
If this duration exceeds $T_P$, the time by which the next packet arrives, the queue size becomes infinite and the transmitter is always on. Otherwise, the probability of the transmitter to be active ($\Phi(\rho)$) is given by the ratio of $T_d$ to $T_P$. Thus we have:
\begin{equation}\label{fi}
\Phi(\rho) = \max\left(\frac{R_p }{R_a(\rho)},1\right)
\end{equation}

In this work, we focus on an OFDMA system that suits LTE standards, and obtain the throughput $R(\rho)$ by link level simulations as described in \cite{{vtc-orange}}. The values taken for $R(\rho)$ from \cite{vtc-orange}, are in fact, averaged over the fast fading and are thus suitable for our model. When there are several users in the network, the available bandwidth is divided among the active users. We assume the bandwidth allocation to be equal among all users and this implies that if $N$ users are all active and experience the same radio conditions, the throughput is reduced to $\frac{R(\rho)}{N}$.

\subsection{Proposed performance metric}

In the broadcast channel there are multiple users that have to be served. In practice, users arrive randomly, and depart once they finish downloading their requested data. New arrivals are blocked when the total number of users crosses a certain limit defined by the base station. Each user may experience a different radio condition from its peers.

For convenience, we divide the area covered by the base station into ``zones". Every user in the same zone, experiences the same radio conditions. This implies that if the base station transmits at a certain power, then all the users in the same zone experience the same SINR. The radio conditions are determined by the average distance of the zone to the base station. If we have $M$ zones in total, we can define $\{\sigma_1^2,\sigma_2^2,\cdots,\sigma_M^2\}$ as the channel conditions for each zone. We then define the ``state" of the system $\sv = \{N_1,N_2,\cdots,N_M\}$. The state $\sv$ represents the number of users in each zone. For example if there are two zones, and there are no users the state is $\{0,0\}$. When a user arrives to zone one, the state becomes $\{1,0\}$.

For a state  $\sv =\{N_1,N_2,\cdots,N_M\}$, the power allocation scheme defined as $\Pv = \{P_1,P_2,\cdots,P_M\}$ results in an SINR distribution of $\rhov=\{\rho_1,\rho_2,\cdots\rho_M\}$ among the zones $1$ to $M$, where $\rho_j=\frac{P_j}{\sigma^2_j}$.

First, we define the notion of energy-efficiency for a given state or the ``local" energy-efficiency. This is useful as in practice, the base station can easily measure this quantity only for a given state as it is unable to predict when a new user will arrive.
The ``global" energy-efficiency defined as the average of the energy-efficiency in each state weighted by their probabilities.

If there is always one and only one user, the energy-efficiency can be defined based on \cite{goodman} and other works as
\begin{equation}
\eta_{SU}= \frac{ R(\rho) \Phi(\rho)}{b  + P \Phi(\rho) }
\end{equation}
where $b$ is the constant power consumed by the base station while serving at least one user\footnote{This cost can have several origins like energy spent on the power amplifier, computation, cooling mechanisms etc. Details of the power consumption model are given in \cite{salah}.}. The proposed form is easy to interpret as $R(\rho)$ represents the average throughput when the transmitter is active and $P$ is the cost when the transmitter is active.

When the system is state $\sv$, the energy-efficiency is defined as:
\begin{equation}
\eta_{\sv}(\Pv)= \frac{\bar{R}_{\sv}(\rhov)} {\bar{P}_{\sv}(\Pv)}
\end{equation}
where $\bar{R}_{\sv}$ and $\bar{P}_{\sv}$ represent the total throughput and power consumed respectively in state $\sv$.

When the number of users is random, then the global energy-efficiency function is defined as:
\begin{equation}
\hat{\eta}= \sum_{\sv} \frac{ \pi({\sv}) \bar{R}_{\sv}  }{\bar{P}_{\sv}}
\end{equation}
Where $\pi({\sv})$ is the probability of finding the base station at state ${\sv}$ of user distribution. The global energy-efficiency could alternately be defined as ratio of the total throughput over all states to the total power over all states. However, in practice, calculating the energy-efficiency for each state and taking the average, is easier and more reasonable. The goal of this work is to improve the above defined energy-efficiency of a transmitting base station.

This metric can be physically interpreted as the average number of bits that can be transmitted by spending one Joule of energy. Alternately, the average power cost of the base station can be written an $\frac{\mathrm{Traffic}}{\eta}$. Hence, optimizing the global energy-efficiency amounts to minimizing the average power consumption of the base station.

\section{Optimal power allocation for a fixed number of users}

In this section we consider the case where the number of users is fixed. We will refer to the optimization of the metric defined in this section as ``local" optimization as it deals with the optimization of a single state of the wireless network. When the state of the network is given, we know the number of users in each zone and can thus calculate the relevant information required to obtain and optimize the energy-efficiency. For our calculations we assume a knowledge of the average noise levels for each zone, i.e $\{\sigma_1^2,\sigma_2^2,\cdots,\sigma_M^2\}$ are known.

\subsection{Homogeneous radio conditions}

First, we consider the problem where all users experience the same average SINR, as the model is easier to be understood; the case of heterogeneous SINRs will be exposed next. Let the total number of users in the cell be $N$. As all the users experience the same radio conditions, $\sv = \{N\}$. In this case if we define the average throughput experienced by any queue as $R_a$, we can derive:
\begin{equation}
R_{a}(\rho) = \sum_{i=0}^{N-1} \binom{N-1}{i}\Phi(\rho)^i (1-\Phi(\rho))^{N-1-i} \frac{R(\rho)}{i+1}
\label{eq:sz:st}
\end{equation}
where $\Phi(\rho)$ denotes the probability that any of the $N$ users are actively being served and is given as in equation \ref{fi}. The summation is upto $N-1$ as $R_a$ is the throughput experienced by an active user, and so we consider the remaining $N-1$ users. The $R_a$ for every user is identical as all users experience the same SINR for the same transmit power. This symmetry can be exploited to conclude that the transmit power for each user will be equal when optimized. Note that $R_a(\rho)$ depends on $\Phi(\rho)$ and $\Phi(\rho)$ depends on $R_a(\rho)$ leading to a fixed point equation.

Clearly if $N$ is large enough, then the demand in data rate will exceed the maximum available throughput and $\Phi(\rho)$ becomes $1$. On the other hand, if $N$ is small enough, the users may transmit their data faster than the packet arrival speed causing the queue to empty occasionally. In this period, other users can take advantage of the excess bandwidth. 

From $\Phi(\rho)$, the total power consumed can be calculated as
\begin{equation}
\bar{P}_{\sv} = b + P (1-(1-\Phi(\rho))^N)
\label{eq:sz:pwr}
\end{equation}
Here $(1-\Phi(\rho))^N$ is the probability of all queues being empty. If any queue is active the power consumed is $P$. The total throughput is $\bar{R}_{\sv}=N \Phi(\rho) R_{a} $ leading to an energy-efficiency of
\begin{equation}
\eta_{\sv} = \frac{ N \Phi(\rho) R_{a}(\rho) }{ b + P (1-\Phi(\rho))^N}
\label{eq:sz:eta}
\end{equation}

\subsection{Heterogeneous radio conditions}

Consider a more realistic setting where users experience different radio conditions in each zone. Denoting the average throughput experienced by zone $j$ as $R_{a:j}$, we can compute
\begin{align}
&R_{a:j}(\rhov) =  R(\rho_j) \sum _{i_1=0}^{N_1} \sum _{i_2=0}^{N_2} \cdots \sum_{i_j=0}^{N_j-1}
\cdots \sum_{i_M=0}^{N_M}  \binom{N_1}{i_i} \nonumber  \\ &\times \binom{N_2}{i_2} \cdots \times \binom{N_j-1}{i_j}\times\cdots  \times\binom{N_M}{i_M}   \times(\Phi_1(\rhov))^{i_1} \nonumber \\ & \times (\Phi_2(\rhov))^{i_2} \times \cdots \times (\Phi_M(\rhov))^{i_M}\times (1-\Phi_1(\rhov))^{N_1-i_1} \nonumber \\& \times (1-\Phi_2(\rhov))^{N_2-i_2} \times
\cdots \times(1-\Phi_j(\rhov))^{N_j-i_j-1} \nonumber \\&\times \cdots \times(1-\Phi_M(\rhov))^{N_M-i_M}  \times \frac{1}{i_1+i_2+ \cdots +i_M+1}
\label{eq:mz:st}
\end{align}
where
\begin{equation}
\Phi(\rhov)_j = \max\left(\frac{R_p}{R_{a:j}(\rhov)},1\right)
\label{eq:mz:phi}
\end{equation}

Leading to a set of fixed point equations that can be solved to calculate all $R_{a:j}(\rhov)$ for a given $\Pv$. Equation (\ref{eq:mz:st}) is similar to (\ref{eq:sz:st}), but considers the presence of users in other zones as well. The average power can be calculated as
\begin{align}
& \bar{P}_{\sv}(\Pv) =   b + \sum_{i_1=0}^{N_1}\cdots  \sum_{i_M=0}^{N_M} \left(1-\delta(i_1+\cdots + i_M)\right)\nonumber \\& \times (\Phi_1(\rhov))^{i_1}\times \cdots \times (\Phi_M(\rhov))^{i_M} \times \frac{ P_1 i_1 +\cdots +P_M i_M }{ i_1+\cdots +i_M } \nonumber \\& \times (1-\Phi_1(\rhov))^{N_1-i_1} \times \cdots \times (1-\Phi_M(\rhov))^{N_M-i_M}
\label{eq:mz:pwr}
\end{align}
Where the $\delta$ function is used to exclude the state where all zones are empty ($\delta(x)=0$ for all real $x$ but $0$, and $\delta(0)=1$). The energy-efficiency in this state can be calculated with $\bar{R}_{\sv}(\rhov) = \sum_{i=1}^M N_i \Phi(\rhov)_i R_{a:i}$ and total power from equation (\ref{eq:mz:pwr}).

\section{Optimal power allocation considering the dynamic behavior of users}

In the previous section, we optimized the energy-efficiency for fixed numbers of users. To analyze the impact of power allocation on the network performance and account for the users arrivals and departures, a flow-level capacity analysis is required. The arrival rate can be modeled through a Poisson process (of intensity $\lambda_i$ in zone $i$) and users leave when they finish streaming a file of average size $F$ (we assume that $F$ is the same for all users). When the total number of users exceed a given threshold $N_{max}$, new user arrivals are blocked.

\subsection{Processor sharing analysis}

When users with a finite workload are considered, the number of users is not constant but varies dynamically during time. The distribution of the number of users is determined by the traffic intensity within the cell. Indeed, if the traffic intensity is large, more users connect to the system per unit time and the average number of active users increases. In this section, we show how to compute the distribution of the number of users knowing the traffic intensity.

The heterogeneity in radio conditions translates into a larger service time for cell
edge users. When the system is in state $\sv = \{N_1,N_2,\cdots,N_M\}$, the total number of users in the cell is $N(\sv)=N_1+\cdots+N_M$. Based on \cite{richard}, we can model the system as a Generalized Processor Sharing queue, whose evolution is just described by the overall number of users in a cell. The solution of the Markov process has the
simple form
\begin{equation}
\pi(\sv) = \frac{1}{\Gamma} \frac{N(\sv)!} {\prod_{i=1}^M N_i!} \prod_{c=1}^M \frac{\Omega_c^{N_c}}{ \prod_{j=1}^{N_c}  j \Phi_{c;\sv(N_c=j)}  R_{a:c;\sv(N_c=j)} }
\end{equation}
where $\Omega_c = S \lambda_c $ and $\Gamma$ is a normalizing constant. The notation $\sv(N_c=j)$ is used to take the $\Phi$ and $R_a$ for the state $\sv$ with $j$ users in zone $c$.

In this model, the user blocking rate can be calculated as $\alpha = \sum_i \lambda_i \sum_x \pi(x)$, $x$ such that the system is full ($N(x)=N_{max}$). Quality of service (QoS) is measured through the user blocking rate. The QoS constraint is thus $\alpha \leq \epsilon$, where $\epsilon$ is the maximum tolerable blocking rate.

\subsection{Optimal power allocation}

The steady-state probabilities defined in the previous section are calculated knowing the throughputs for each state of the network. This throughput will of course depend on the power allocation as explained in Sections II and III. The power allocation has thus to be optimized taking into account the dynamics of users.  A power allocation policy $\hat{\Pv}$ is defined as a set of actions for each of the possible states:

\begin{equation}
\hat{\Pv}=\bigcup_{\sv}\Pv_{\sv}
\end{equation}

The global energy efficiency; knowing the policy $\hat{\Pv}$, is given by:
\begin{equation}
\hat{\eta}(\hat{\Pv}) = \sum_{\sv} \frac{ \pi({\sv}) \bar{R}_{\sv}(\rhov) }{\bar{P}_{\sv}(\Pv_{\sv}) }
\end{equation}
The optimization problem can be defined as
\begin{equation}
\hat{\Pv^*}= \arg\max[ \hat{\eta}(\hat{\Pv}) ]
\end{equation} 
And the maximum global energy-efficiency possible is $\hat{\eta}(\hat{\Pv^*})$.

The idea behind this global optimization is that the power allocation does not depend uniquely on the actual state of the network, but takes also into account the future evolutions of the network. For instance, a power allocation decision that is taken at one moment may have an influence on the evolution of the state of the network by favoring a subset of users by a better throughput. We will study in the next section the difference between this global policy maximization and a local one, as defined in section III.

\section{Numerical results}

In this section, we use simulations and numerical calculations to study the properties of the energy-efficiency function and obtain the power allocation that maximizes it. We consider the receiver and the transmitter to have two antennas each forming a $2 \times 2$ MIMO system. The data rates for this configuration which are LTE compliant are taken from \cite{vtc-orange} and are given as a function of the SINR. For the single zone case we take $\sigma^2=1$ mW while for the two zone case we have $\{\sigma_1^2,\sigma_2^2\}=\{1,\frac{1}{8}\}$ mW. We begin by illustrating the results when the network is optimized supposing that the number of users is fixed. The dynamic behavior of users is taken into account afterwards and the performance of the network is compared for both schemes.

\subsection {Numerical results for the local optimization}

\begin{figure}[H]
    \begin{center}
        \includegraphics[width=100mm]{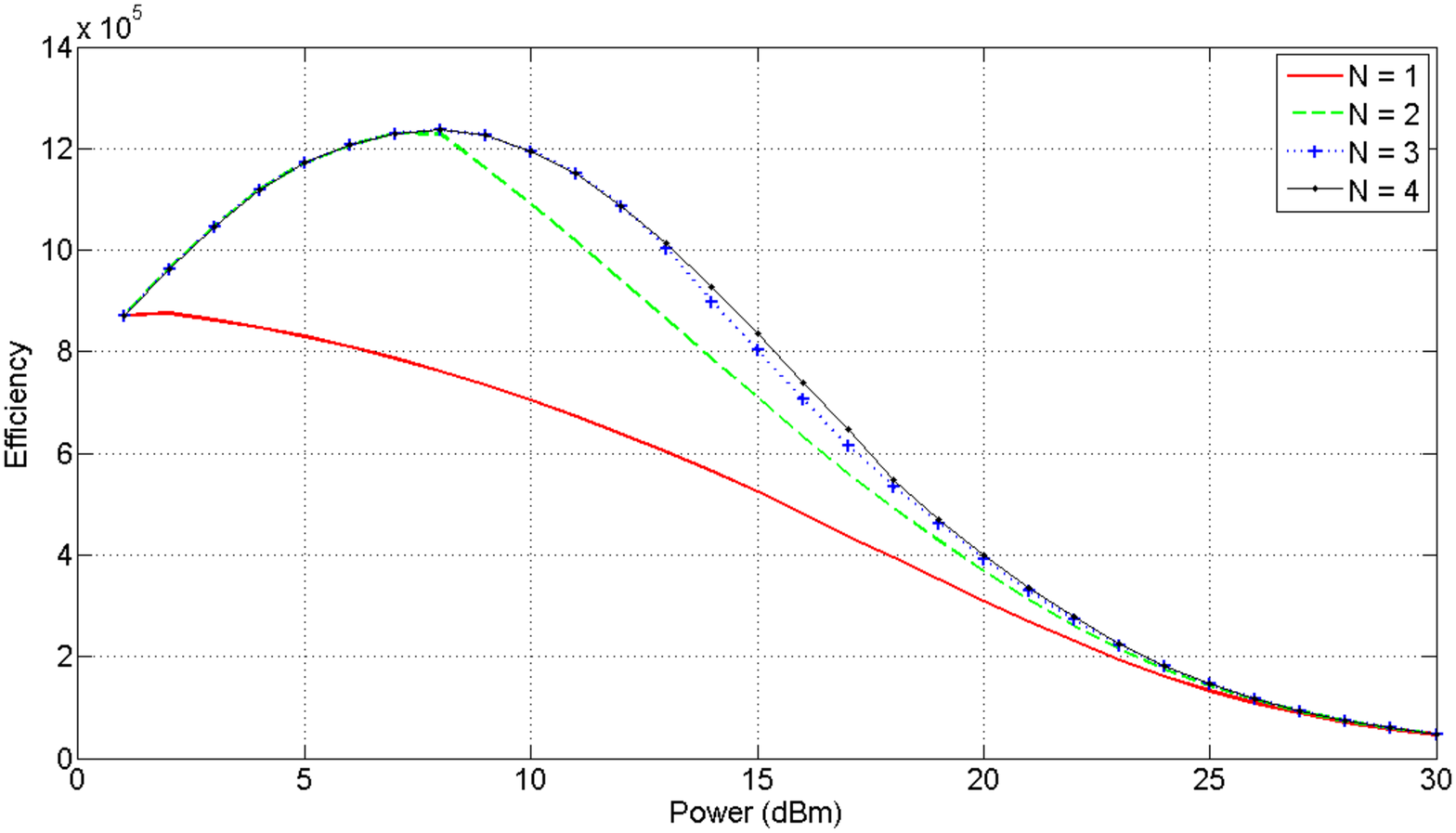}
    \end{center}
    \caption{$\eta$ vs $P$ with $\frac{b}{\sigma^2}=100$ (20dB). Note that the energy-efficiency is peaked at higher powers with additional users. }
    \label{fig:eest}
\end{figure}

We begin by illustrating the power allocation scheme when the dynamic behavior of users is not taken into account, and when all users are subject to the same radio conditions. In figure \ref{fig:eest}, we show the energy-efficiency as a function of the transmit power. Here, due to symmetry, all the users use the same power. The results show that the energy efficiency begins by increasing with the transmit power increases, as users are able to reach higher throughputs. However, starting from one point, users reach the maximal throughput they are able to reach as, in LTE, modulation schemes are limited; the energy efficiency begins thus decreasing as throughputs remain constant while power consumption increases.

\begin{figure}[h]
    \begin{center}
        \includegraphics[width=100mm]{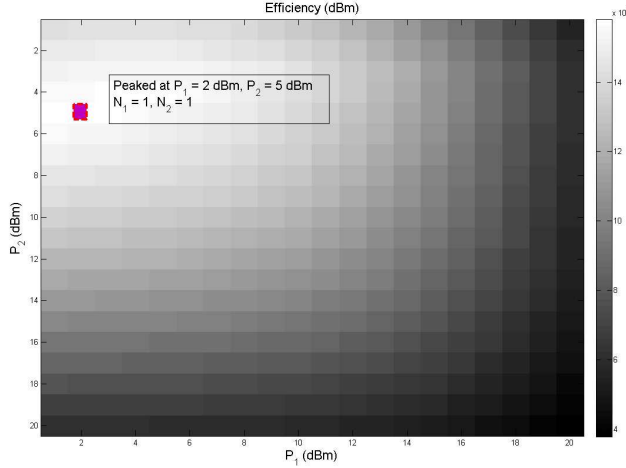}
    \end{center}
    \caption{$\eta$ over combinations of $P_1$ and $P_2$ with $\frac{b}{\sigma_1^2}=100$ (20dB), $N_1=N_2=1$. Zone $2$ corresponds to a lower SINR and in this case the efficiency is optimized by using more power on the zone $2$ user. }
    \label{fig:ee2z11}
\end{figure}

\begin{figure}[h]
    \begin{center}
        \includegraphics[width=100mm]{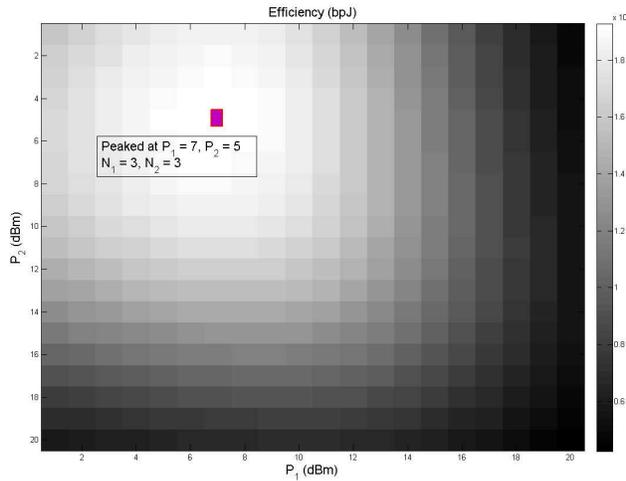}
    \end{center}
    \caption{$\eta$ over combinations of $P_1$ and $P_2$ with $\frac{b}{\sigma^2}=100$ (20dB), $N_1=N_2=3$. As before, zone $2$ corresponds to a lower SINR and interestingly, in this case, the efficiency is optimized by using more power along the zone $1$ user. This is because with 3 users in each zone, the demanded rate exceeds the maximum available throughput and so, optimization is done by using power on users with a better SINR.}
    \label{fig:ee2z33}
\end{figure}

In figure \ref{fig:ee2z11}, we consider the case of two users: one in the ``inner" zone (near base station) and the other in the ``outer" zone (at cell edge). In this case, the system has a sufficient capacity to support both users and the energy efficiency is optimized when more power is used on the outer zone which compensates for its lower SINR. Here the total throughput can thus be increased by using more power on the outer zone user. However in figure \ref{fig:ee2z33}, we have three users in both the inner and outer zones. Here the throughput of the wireless network is not sufficient for all the users and so the energy-efficiency is optimized by simply putting more power in the inner zone with the higher SINR as the total throughput is not improved by putting more power into the outer zone.


\subsection{Numerical results for the global optimization}
We have illustrated, till now, the performance of the system when the number of users is fixed. In this section, we consider the dynamic behavior of users. In this setting, the power allocation is not determined for a fixed number of users, but for a given traffic intensity. the number of users is thus a random variable whose distribution depends on the traffic intensity. The optimal power allocation is the one that maximizes the energy efficiency while maintaining a constraint on the QoS. Note that this optimal power allocation is a matrix that gives, for each state of the network composed of the number of users in the cell, the power allocation for each of the users. 

\begin{figure}[h]
    \begin{center}
        \includegraphics[width=100mm]{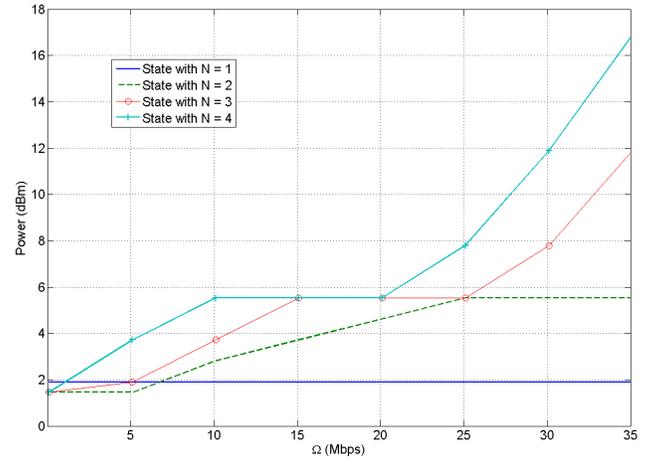}
    \end{center}
    \caption{The power allocation scheme $(P_1,\cdots,P_4)$ plotted against the traffic $\Omega$ when $\hat{\eta}$ is optimized. Also note that the QoS constraint of maintaining the blocking rate below $0.01$ is satisfied.}
    \label{fig:gpa}
\end{figure}

Initially we consider the cell with homogeneous radio conditions, i.e. we suppose that all the users experience the same SINR on average. In this setting, if $N_{max}$ is the maximum number of users allowed, optimization is performed over $N_{max}$ variables, i.e. the power used in each state. For the single zone case we take $\sigma^2=1$ mW. 
The optimal power allocation is shown in figure \ref{fig:gpa}. Note that, in this case, the power allocation is a vector and not a matrix, as all users experience the same radio conditions and have, by symmetry, the same allocated power.

\begin{figure}[h]
    \begin{center}
        \includegraphics[width=100mm]{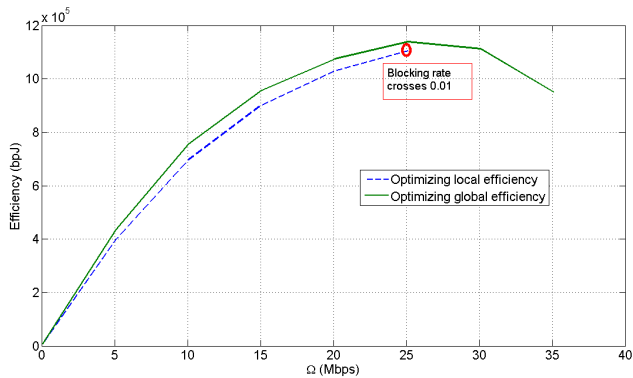}
    \end{center}
    \caption{$\hat{\eta}$ plotted against the traffic $\Omega$ when $\hat{\eta}$ is optimized and when $\eta$ is optimized for each state separately.}
    \label{fig:g1z}
\end{figure}

\begin{figure}[h]
    \begin{center}
        \includegraphics[width=100mm]{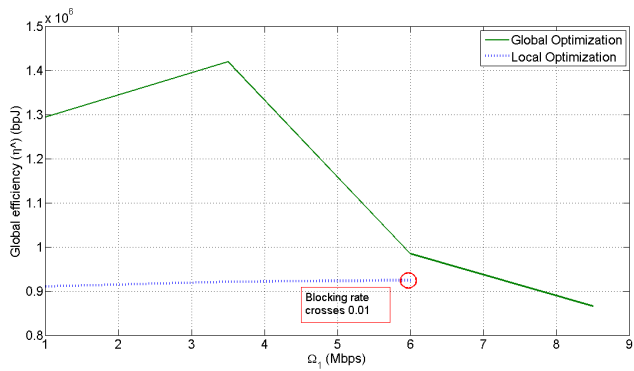}
    \end{center}
    \caption{$\hat{\eta}$ plotted against the traffic $\Omega_1 = \lambda_1 S$ when $\hat{\eta}$ is optimized and when $\eta$ is optimized for each state separately. Also note that the QoS constraint of maintaining the blocking rate below $0.01$ is satisfied at all points shown.}
    \label{fig:g2z}
\end{figure}

Figure \ref{fig:g1z} compares the energy-efficiency obtained for the local and the global optimizations. Recall that, by local, we mean that the optimization is done for each state independently from the others, taking into account only the observed number of users and not the future evolutions of the system. As seen from the simulations (Figure \ref{fig:g1z}), using a global optimization does not seem to yield much gains in the energy-efficiency for the single zone case. This is because the throughput, and thus service times, are the same for all users. We next move on to the two-zone case (cell center and cell edge). Here we consider a cell divided into two concentric rings, and define the outer zone as the region when the SINR is $4.8$ dB (3 times) lower than the SINR for the inner zone, when the transmit power is unchanged. The outer zone also has $3$ times the area of the inner zone causing $\lambda_2 = 3 \lambda_1$. With these parameters we attempt to calculate the optimal global energy-efficiency and corresponding power allocation for given values of $\lambda_1$. We have $\{\sigma_1^2,\sigma_2^2\}=\{1,\frac{1}{3}\}$ mW. Figure \ref{fig:g2z} shows the energy efficiencies corresponding to local and global optimizations. It is obvious that global optimization yields much higher efficiency when users have heterogeneous radio conditions. This is because, in the local optimization setting, the notion of call duration cannot be taken into account as users are considered as always active. The optimal power allocation will then tend to favor cell center users in order to maximize throughput. However, when the dynamic behavior of users is taken into consideration, it is sometimes better to use more power on cell edge users in order to let them finish their service quickly and quit the system. Applying the policy obtained from the local optimization will lead to users accumulating at the cell edge as they are not able to finish their transfers.

\section{Conclusion}

In this work we study and optimize the flow level energy efficiency of base stations in LTE. We introduce the notion of a ``global" energy-efficiency which is defined as the average of the energy-efficiencies of each state the cell can be in. These states represent the traffic configurations, i.e. the numbers and positions of users in the cell. Through extensive simulations we see that optimizing the global efficiency yields a different power allocation from optimizing the efficiency of each individual state. Although this difference can be neglected when considering a cell in which all users experience the same average SINR, when considering a more realistic setting where users are subject to heterogeneous radio conditions, the global optimization yields a considerable gain. This is because, when users are considered as static, it may be optimal to give more power to cell center in order to increase throughputs. However, when the dynamic behavior of users is taken into account, giving more power to users with bad radio conditions will allow them leaving the system faster and thus alleviating load in the future. When compared to the local optimization, it is observed that the global optimization improves the energy-efficiency up to a factor of 50\%.

\section*{Acknowledgments}

This work is a joint collaboration between  Orange Labs, Laboratoire des signaux et syst\'{e}mes (L2S) of Sup\'{e}lec and the Alcatel Lucent Chair of Sup\'{e}lec. This work is part of the European Celtic project ``Operanet2".

\end{document}